\documentclass[pss]{wiley2sp} % provides pss two-column style
\usepackage{amsmath}
\newcommand{\wt}{\omega_{\mathrm{t}}}
\newcommand{\wh}{\omega_{\mathrm{h}}}

\begin{document}

% Title of the article
\title{Nonlinear optical response of hole--trion systems in quantum
dots in tilted magnetic fields}

% Abbreviated title for the page headers
\titlerunning{Nonlinear optical response of hole--trion systems}

% Authors
\author{%
  Pawe{\l} Machnikowski\textsuperscript{\Ast,\textsf{\bfseries 1}},
  Tilmann Kuhn\textsuperscript{\textsf{\bfseries 2}}}

% Abbreviated list of authors for the page headers
\authorrunning{P. Machnikowski, T. Kuhn}

%E-mail-address of corresponding author
\mail{e-mail
  \textsf{Pawel.Machnikowski@pwr.wroc.pl}, Phone:
  +48-71-3204546, Fax: +48-71-3283696}

% author's affiliations/addresses
\institute{%
  \textsuperscript{1}\,Institute of Physics, Wroc{\l}aw University of
Technology, 50-370 Wroc{\l}aw, Poland\\
  \textsuperscript{2}\,Institute of Solid State Theory, 
University of M\"unster, 48149 M\"unster, Germany}

\received{XXXX, revised XXXX, accepted XXXX} % do not change, will be filled in by the publisher
\published{XXXX} % do not change, will be filled in by the publisher

% Please select about four verbal keywords for your manuscript.
\keywords{nonlinear optical spectroscopy, spin, decoherence,
  magnetooptical Kerr effect}

\abstract{%
% This is a macro for the typesetting of two-column text in an
% abstract. It will typeset the two arguments in \abstcol{}{} as the
% left and right column inside the abstract box. At the
% columnbreak there will be always a columnbreak (\par), so both
% columns start with a new paragraph. No automatic column height
% balancing is done.
%
% If used with a \titlefigure it will silently output both
% parameters as consecutive paragraphs.
%
% The macro is defined exclusively inside the argument of \abstract{};
% if used outside it will raise an error.
%
% Usage: \abstcol{<left column>}{<right column>}
\abstcol{%
We discuss, from a theoretical point of view, the four wave mixing
spectroscopy on an ensemble of 
p-doped quantum dots in a magnetic field slightly tilted from the
in-plane configuration. We describe the system evolution in the
density matrix formalism.
In the limit of coherent ultrafast optical driving, we obtain
analytical formulas for the single system dynamics and for the
response of an inhomogeneously broadened ensemble. }{%
The results are compared to the previously studied time-resolved Kerr
rotation spectroscopy on the same system.
We show that the Kerr rotation and four wave mixing spectra yield complementary
information on the spin dynamics (precession and damping).
}}

% The class file requires the standard graphicx Latex package. See the 'LaTeX
% standard graphics and color packages documentation' for more information at
% <http://tug.ctan.org/tex-archive/macros/latex/required/graphics/grfguide.pdf>.
%
% Accepted figure file formats depend on which LaTeX flavour is used.
% Classic LaTeX is always able to use Encapsulated Postscript (EPS);
% PDFLaTeX can't use this but accepts PDF, JPG, PNG, and GIF formats.
%
% See examples for implementing graphics in floating figure environments later in this file.
% If \titlefigure is given, it takes as its mandatory parameter the
% name (without extension) of some figure file.

\maketitle   % please do not remove

\section{Introduction.}

The properties of confined spins in semiconductor nanostructures are
interesting because of their expected important role in quantum computing
and spintronic devices. In particular, the hole spin attracts much
attention because of its enhanced coherence time. In systems that
efficiently couple to optical 
fields, like quantum wells and self-assembled quantum dots, an
interesting possibility is to study the spin dynamics using optical
spectroscopy tools. A method which is particularly suited for this
purpose is the time-resolved Kerr (or Faraday) rotation (TRKR, TRFR)
\cite{syperek07,kugler09,korn10}, where the evolution of the occupations of the hole
and trion Zeeman states is traced by investigating the rotation of the
polarization plane of the reflected or transmitted probe pulse.

In a recent work
\cite{machnikowski10}, we proposed a general description for the
dynamics of a confined hole-trion system in a tilted magnetic
field. We showed that a TRKR experiment provides rich
information about the rates of spin precession and decoherence (both
longitudinal and transverse). In particular, the optical response at
slightly tilted magnetic fields contains contributions related both to
the hole spin relaxation (longitudinal decoherence) with respect
to the hole spin quantization axis, as well as the dephasing of spin
coherences (transverse decoherence) with respect to this axis. Under
favorable experimental conditions (a sufficient separation of time scales), the
corresponding two decoherence rates can be deduced from a single run
of an experiment, although in a realistic system the latter may be
convoluted with (or even dominated by) the inhomogeneous dephasing.

Although the TRKR or TRFR method may be the most obvious choice for the
investigation of the spin dynamics the spin precession
of the trion and hole in a tilted field will lead to transitions between optically active
and inactive states which should be manifested in any form of the
optical response. In particular, signatures of the spin-related
dynamics should be visible in the four wave mixing (FWM) nonlinear
spectroscopy. As the FWM spectroscopy is one of the most widely used
methods in the investigation of semiconductors and their
nanostructures \cite{borri01,vagov04}
it may
be interesting to study how the spin precession and decoherence affect
the FWM response and what information on the spin-related kinetics can
be extracted from this kind of experiments.

In this contribution, we study the FWM response of the same system of
confined holes as discussed in our previous work
\cite{machnikowski10}. We show  
that the FWM signal is also affected by the spin dynamics in a way
that allows one to extract the Larmor frequencies of electrons and
holes. The available information on the decoherence rates
is less specific than that contained in the TRKR response and is
available only as long as the hole spin dephasing times are not much longer
than the life time of the optical coherence. On the other hand, the
spins undergo partial refocusing in the two-pulse optical echo
experiment which leads to the appearance of components in the optical
response that depend on the inherent spin dephasing rates but are
insensitive to the inhomogeneous distribution of the Larmor
frequencies. In this way, the FWM method may be a valuable
experimental tool to study the spin dynamics, complementary
to the TRKR or TRFR techniques. 

The paper is organized as follows. In Sec.~\ref{sec:model}, we define
the model under study. In Sec. \ref{sec:evol} we derive the FWM
response of the system. Next, in Sec.~\ref{sec:discusion}, we discuss
the result and compare the spin-related information contained in the
FWM signal to that available from the TRKR response.

\section{Model.}
\label{sec:model}

We study a system similar to that discussed in our previous work
\cite{machnikowski10}, consisting of an ensemble of quantum dots or
trapping centers in quantum wells with one confined hole in each of the
dots. The system is excited with a pump pulse at the time $t=-\tau$ and a
probe pulse at $t=0$. Unlike in our previous work, we are now
interested in the FWM response from the system. Experimentally, the
relevant third-order response is isolated by choosing the
appropriate excitation and detection directions (see
Ref.~\cite{machnikowski10} for a discussion). In the modeling, we
assume the pump and probe pulses to have phases $\phi_{1}$ and
$\phi_{2}$, respectively, and calculate the terms in the response of an
inhomogeneously broadened system that carry the phase
$2\phi_{2}-\phi_{1}$. Out of many possible configurations of the
polarizations of the excitation and detection, we choose the
circularly co-polarized one, with both pulses having the $\sigma_{+}$
polarizations and the detection being performed at the same
polarization. 
The system is placed in a magnetic field
tilted by the angle $\vartheta$ to the normal, which defines the trion spin
quantization. The hole spin is quantized
along the axis at an angle $\varphi$ from the normal to the
sample, defined by the components of the highly anisotropic hole
Land\'e tensor. We will assume that the system geometry is very close
to the Voigt configuration, with the magnetic field only slightly
tilted from the system plane, so that $\cos\varphi\approx 0$ and
$\sin\varphi\approx 1$.

The system dynamics is a composition of many processes. For the optical
hole-trion transition, we assume the radiative recombination rate
$\gamma_{1}$ and the additional pure dephasing rate $\gamma_{0}$. The
hole and the trion spin precess in the magnetic field with the Larmor frequencies
$\wh$ and $\wt$, respectively. The longitudinal relaxation rate for
the hole spin is
$T_{1}^{-1}=\kappa_{+}+\kappa_{-}$, where $\kappa_{-},\kappa_{+}$ are the rates
for the spin flip transitions to the lower and to the upper Zeeman
level, related by the detailed balance condition
$\kappa_{-}/\kappa_{+}=\exp(\hbar\wh/k_{\mathrm{B}}T)$, where
$k_{\mathrm{B}}$ is the Boltzmann constant and $T$ is the
temperature. The hole spin transverse relaxation rate is
$T_{2}^{-1}=T_{1}^{-1}/2+\kappa_{0}$, where $\kappa_{0}$ is the
additional pure dephasing rate. We assume that the electron spin dephasing is slow
compared to the trion lifetime so that the electron spin coherence is
dominated by the latter. 

The simulations presented in this contribution are performed for
$T=4$~K and for the magnetic field of 7~T tilted at an angle
$\pi/2-\vartheta=4^{\circ}$ from the system plane. For the in-plane and
perpendicular components of the hole g-tensor equal to $0.04$ and
$0.6$, respectively, this yields the hole Larmor frequency $\wh=0.036$~ps$^{-1}$ and
the hole spin quantization axis oriented at $\varphi=44^{\circ}$ from the
normal direction. The trion Larmor frequency is 0.16~ps$^{-1}$
(corresponding to the electron g-factor of 0.26). We will assume the
decoherence times $1/\gamma_{1}=1.2$~ns and $T_{1}=2$~ns and no
additional pure dephasing effects ($\gamma_{0}=\kappa_{0}=0$).

\section{The FWM response.}
\label{sec:evol}

Our theoretical analysis is based on the method developed in
Ref.~\cite{machnikowski10}: the system evolution is studied in the
density matrix formalism, with the optical and spin-related dephasing
included via a Lindblad dissipator in the evolution equation. In the
limit of coherent ultrafast optical driving, this approach yields
analytical formulas for the single system dynamics, which allows one to
perform averaging over an inhomogeneous distribution of various
parameters in the ensemble. 

For a $\sigma_{+}$ probe, the only element of the density matrix 
just after the probe pulse that carries the $e^{2i\phi_{2}}$ phase
dependence is 
\begin{equation*}
\rho_{31}(0^{+})=\rho_{13}(0^{-})e^{2i\phi_{2}}
\sin^{2}\frac{\alpha_{2}}{2},
\end{equation*}
where $\phi_{2}$ is the phase of the probe pulse, $\alpha_{2}$ is its
area, and $0^{-}$ denotes the time instant just before the arrival of
the pulse. The evolution of the system state is then calculated using
the Master equation in the Lindblad form, like in Ref.~\cite{machnikowski10}. The
$\sigma_{+}$ interband coherence at a time $t>0$ is
\begin{equation*}
\rho_{31}(t)=\sum_{\pm,\pm}
d_{\pm,\pm}e^{\lambda_{\pm,\pm}t}\rho_{13}(0^{+})
e^{-iEt/\hbar},
\end{equation*}
where the amplitudes and the exponents are given by
\begin{eqnarray*}
d_{1}=\frac{\cos\varphi+ 1}{4}, & \quad &
\lambda_{1}=\frac{2i\wt -(2i\wh -\beta)}{4}, \\
d_{2}=\frac{\cos\varphi- 1}{4}, & \quad &
\lambda_{2}=\frac{2i\wt +(2i\wh -\beta)}{4}, \\
d_{3}=-\frac{\cos\varphi+ 1}{4}, & \quad &
\lambda_{3}=\frac{-2i\wt -(2i\wh -\beta)}{4}, \\
d_{4}=-\frac{\cos\varphi- 1}{4}, & \quad &
\lambda_{4}=\frac{-2i\wt +(2i\wh -\beta)}{4}. \\
\end{eqnarray*}
Here $E$ is the interband transition energy (at zero field) and
the dephasing constants are
\begin{equation*}
\Gamma=4\gamma_{0}+2\gamma_{1}+\kappa_{0}+\kappa_{+}+\kappa_{-}
\end{equation*}
and 
\begin{equation*}
\beta=(\kappa_{-}-\kappa_{+})\left[ 
(\cos\varphi-1)\sin^{2}\varphi +1\right].
\end{equation*}

Now, we need to find $\rho_{13}(0^{-})$. For a $\sigma_{+}$--polarized
pump, just after the pump pulse (at $t=-\tau^{+}$),
the only non-zero element linear in the pump amplitude is
$\rho_{31}$. By
solving the Master equation one finds at $t=0^{-}$
\begin{equation}
\label{ro13}
\rho_{13}(0^{-})=\sum_{i}
c_{i}e^{\lambda_{i}^{*}\tau}e^{iE\tau/\hbar},
\end{equation}
where 
\begin{equation*}
c_{i}=\rho_{13}(-\tau^{+})d_{i}.
\end{equation*}
Thus, at the time $t$ one finds
\begin{eqnarray*}
\rho_{31}(t) & = & \frac{1}{2}\sin\alpha_{1}\sin^{2}\frac{\alpha_{2}}{2}
e^{i(2\phi_{2}-\phi_{1})} \\
&&e^{-iE(t-\tau)/\hbar}
\sum_{ij} d_{i}d_{j}e^{\lambda_{i}t+\lambda_{j}^{*}\tau}.
\end{eqnarray*}
When averaging over the distribution of interband energies $E$ in a
typical QD ensemble, the exponent produces a narrow echo peak around
$t=\tau$. Since the spin-related evolution (frequencies $\lambda_{i}$)
is slow on this time scale, we can put $t=\tau$ under the
summation. Then the magnitude of the time-integrated response is
proportional to
\begin{eqnarray}
\lefteqn{\mathrm{FWM}  \sim  \left|
\sum_{i}d_{i}e^{\lambda_{i}\tau}\right|^{2}} \nonumber \\
&& =  \frac{1}{8}(1+\cos\omega_{\mathrm{t}}\tau)
e^{-\Gamma\tau/2}\Bigl[ 
(\cos\varphi-1)^{2}e^{-\beta\tau/2} \nonumber \\
&&\;\;\;\;+(\cos\varphi+1)^{2}e^{\beta\tau/2}
+2(1-\cos^{2}\varphi)\cos\omega_{\mathrm{h}}\tau \Bigr].
\label{fwm}
\end{eqnarray}
In order to take into account also the inhomogeneity of the Land\'e
tensors at different trapping centers this result should be averaged
over a certain distribution of the Larmor frequencies $\wh$ and $\wt$,
which we will assume to be Gaussian and characterized by the variances
$\sigma_{\mathrm{h}}^{2}$, $\sigma_{\mathrm{t}}^{2}$. We assume here
that the distribution of Land\'e tensors is uncorrelated to the
spectral positions of the trion transitions.

\begin{figure}[t]
\includegraphics*[width=\linewidth]{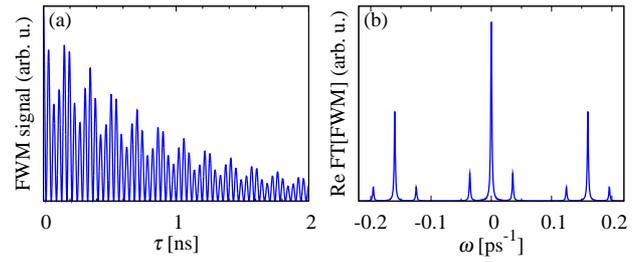}
\caption{The FWM signal from a system without inhomogeneous broadening
  of the Larmor frequencies. (a) The
  time-integrated FWM signal as a function of the delay time; (b) The
  real part of the Fourier transform of the time-resolved signal.}
\label{fig:homo}
\end{figure}

In Fig.~\ref{fig:homo}(a), we show the calculated time-integrated FWM
signal in the absence of inhomogeneous distribution of the Larmor
frequencies. As follows from Eq.~\eqref{fwm}, oscillations at various
frequencies are present in the optical signal, corresponding to
combinations of the hole and trion frequencies. A more transparent
picture is obtained after performing a Fourier-transform of this
signal, as shown in Fig.~\ref{fig:homo}(b). Here one can see the zero-frequency
line, as well as lines at the frequencies $\pm\wt,\pm\wh$, and
$\pm\wt\pm\wh$.  

\begin{figure}[t]
\includegraphics*[width=\linewidth]{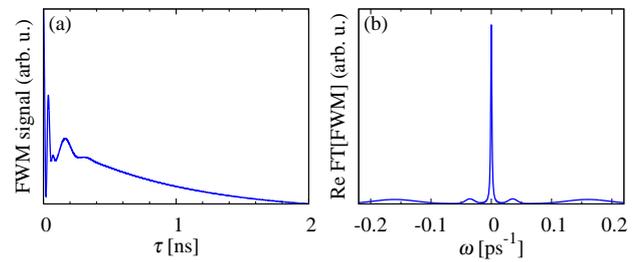}
\caption{The FWM signal from a system with inhomogeneously broadened
  Larmor frequencies. (a) The
  time-integrated FWM signal as a function of the delay time; (b) The
  real part of the Fourier transform of the time-resolved signal.}
\label{fig:inhomo}
\end{figure}

In a real system, the optical response is affected by the inhomogeneous
distribution of the relevant parameters. In the present case, the
Larmor frequencies are usually not identical for each hole-trion
system due to variations of the $g$-factors in the nanostructure. In
Fig.~\ref{fig:inhomo} we show the results of a simulation performed
for the same parameters as in the previous case but with the
additional effect of inhomogeneous broadening of the Larmor frequency,
which is assumed to be equal to 20\% of their average values. Due to
this inhomogeneity effect, the oscillations in the time integrated response
are damped and the signal is dominated by the monotonic decay (except
for short delay times). Correspondingly, all the non-zero frequency
peaks in the Fourier spectrum are strongly broadened. However, the
central peak remains unaffected.

\section{Discussion and conclusions.}
\label{sec:discusion}

An interesting point in the discussion presented above is the presence
of a zero-frequency 
component which produces a central peak in the Fourier transform,
composed of two Lorentzian contributions: one with the width
$(\Gamma+\beta)/2$ and another one with $(\Gamma-\beta)/2$. 
Interestingly, as follows from Eq.~\eqref{ro13}, there are no
zero-frequency components in the polarization evolution between the
pulses. Therefore, the non-oscillatory part of the FWM response can be
interpreted as a result of partial refocusing of the Larmor precession
by the probe pulse.

The presence of this zero-frequency component, which is insensitive to
the inhomogeneous distribution of the $g$-factor, opens a possibility of
extracting useful information on the spin-related system parameters. 
If the trion lifetime is sufficiently long compared to the hole spin
decoherence times ($\beta$ not too small compared to $\Gamma$) then
the two Lorentzian components of the central line, with the widths
$\Gamma\pm\beta$ can be separated by fitting, from which the values of
$\Gamma$ and $\beta$ can be deduced. The value of $\beta$, along with
the detailed balance relation, allows one to extract the rates
$\kappa_{\pm}$, hence the longitudinal decoherence time. On the other
hand, $\Gamma$ involves both the spin-related rates $\kappa_{\pm,0}$
and the optical 
lifetime and dephasing rates $\gamma_{0,1}$. The latter, however, can
often be deduced independently for a given system, which can allow one
to find also the value of $\kappa_{0}$ and to calculate the
intrinsic transverse spin decoherence time $T_{2}$. 

This should be compared with the information available from a TRKR
experiment \cite{machnikowski10}. There, the hole-related response
originates from the occupations of the hole Zeeman sublevels after the
trion recombination. Therefore, the TRKR experiment allows one to
extract the hole spin-related information even if the exciton lifetime
and coherence time are very short, as was indeed the case in the
experiments \cite{syperek07,kugler09}. However, as the information on
the transverse spin dephasing in that experiment is deduced from the
spin precession signal, it is convoluted with the inhomogeneous effect
which may even completely dominate the decay of coherent precession
signal. Therefore, in a TRKR experiment essentially only the
inhomogeneous transverse decoherence time $T_{2}^{*}$ is available.

Thus, we have shown that a FWM experiment performed on an ensemble of
quantum dots or other trapping centers doped with excess holes can provide
useful information on the properties of hole spin precession and
decoherence. Although this information is only available under
favorable conditions related to the various decoherence rates in the
system (long exciton coherence) it is insensitive to inhomogeneous
effects, in particular to a variation of $g$-factors. Therefore, we
conclude that the FWM experiments may be a useful tool to extract
spin-related information which is complementary to that available from
the TRKR study.

\section*{Acknowledgments.} 

This work was supported in part by the
Alexander von Humboldt Foundation within a Research Group Linkage
Grant. P.M. acknowledges support from the TEAM programme of the Foundation for
Polish Science, co-financed from the European Regional Development
Fund.

%\bibliographystyle{pss}
%\bibliography{abbr,quantum}

\providecommand{\WileyBibTextsc}{}
\let\textsc\WileyBibTextsc
\providecommand{\othercit}{}
\providecommand{\jr}[1]{#1}
\providecommand{\etal}{~et~al.}

\end{document}